# Reinforcement Learning for Resource Allocation in Steerable Laser-based Optical Wireless Systems

Abdelrahman S. Elgamal, Osama Z. Alsulami, Ahmad Adnan Qidan, Taisir E.H. El-Gorashi and Jaafar M. H. Elmirghani
*School of Electronics and Electrical Engineering*, *University of Leeds*, Leeds, United Kingdom

*Abstract*— Vertical Cavity Surface Emitting Lasers (VCSELs) have demonstrated suitability for data transmission in indoor optical wireless communication (OWC) systems due to the high modulation bandwidth and low manufacturing cost of these sources. Specifically, resource allocation is one of the major challenges that can affect the performance of multi-user optical wireless systems. In this paper, an optimisation problem is formulated to optimally assign each user to an optical access point (AP) composed of multiple VCSELs within a VCSEL array at a certain time to maximise the signal to interference plus noise ratio (SINR). In this context, a mixed-integer linear programming (MILP) model is introduced to solve this optimisation problem. Despite the optimality of the MILP model, it is considered impractical due to its high complexity, high memory and full system information requirements. Therefore, reinforcement Learning (RL) is considered, which recently has been widely investigated as a practical solution for various optimization problems in cellular networks due to its ability to interact with environments with no previous experience. In particular, a Q-learning (QL) algorithm is investigated to perform resource management in a steerable VCSEL-based OWC systems. The results demonstrate the ability of the QL algorithm to achieve optimal solutions close to the MILP model. Moreover, the adoption of beam steering, using holograms implemented by exploiting liquid crystal devices, results in further enhancement in the performance of the network considered.

## I. Introduction

The spectrum limitations in radio-based wireless communication networks has led researchers towards investigating the use of optical spectrum for new generations of wireless networks. Two types of optical wireless communication (OWC) systems are suitable for deployment in indoor environments: visible light communication (VLC) and infrared-based (IR) communication systems. In VLC systems, high data rates can be achieved using the wide-angle beams of white light emitting diodes (LEDs), which can be used for illumination and data communication simultaneously [1] – [6]. In contrast, infrared (IR)-based wireless systems provide higher data rates which can reach to tens of gigabits per second using the narrow beams of laser diodes (LDs) [7]. Besides, OWC systems can provide higher security in the physical layer compared to radio-based wireless systems due the confined coverage area of optical APs [8], [9].

The current off-the-shelf white LED has low modulation bandwidth (10s of MHz), which might limit the transmission rate that can be achieved using the visible light spectrum. Alternatively, using the narrow beam of vertical cavity surface emitting laser (VCSEL) for data transmission might unlock terabit per second rates in indoor OWC systems due its high modulation bandwidth, in addition to its high efficiency in terms of power conversion. In [10], it has been shown that VCSELs can provide an aggregate data rate that exceeds 1 terabit per second in the indoor environment under eye safety constraints. Furthermore, VCSELs can be implemented for uplink transmissions using various beam steering methods [11], [12].

Several techniques have been used to enhance the performance of OWC systems. Firstly, diversity technologies can improve the signal to interference plus noise ratio (SINR) in [13] – [21]. Secondly, adaptation techniques have been considered to enhance the performance and capacity of downlink channels using beam angle, beam power and beam delay adaptations in [22] – [30]. Thirdly, multiple access techniques such as multi-carrier code division multiple access (MC-CDMA) [22], [30], non-orthogonal multiple access (NOMA) [22] and wavelength division multiple access (WDMA) [33], [34] have been investigated to serve multiple users maximizing the spectral efficiency. Finally, various resource allocation (RA) techniques have been proposed to optimally utilise the resources of the network including wavelength, power, time, and frequency through formulating different optimisation problems [34] - [36].

For the resource management, mixed-integer linear programming (MILP) optimisation can be used to solve different problems with various objective functions providing optimal solutions [34], [35]. However, MILP has two issues that make it unsuitable for many practical scenarios. First, MILP requires full knowledge of the network, which is not easy to provide in practical systems. Second, the time complexity and storage requirements increase with the increase in the dimensions of the network, i.e. the number of transmitters and receivers.

Reinforcement learning (RL) has attracted massive attention for solving such optimisation problems due to its ability to interact with many different systems without any prior knowledge [37]. Basically, RL works on trial and error basis to make decisions within a system aiming to maximise a certain reward [38]. Recently, many studies adopted different RL techniques for solving many optimisation problems with different objectives in a variety of communication networks such as Heterogeneous Cellular Networks (HetNets) [39], Cognitive Radio Networks (CRANs) [40], Mobile Edge Computing (MEC) [41] and Software Defined Networks (SDNs). For OWC systems, the work in [42] introduced a RL-based resource allocation solution for integrated VLC and VLC positioning (VLCP) systems to maximise the sum-rate of users. In [43], time-slots are allocated intelligently using RL in order to maximise the spectral efficiency of VLC systems where a dynamic time-division multiplexing (DTDMA) scheme was considered. The Q-learning algorithm was also adopted in [36] to solve an optimisation problem for resource allocation in a WDM-based VLC system. It is shown that the RL has the ability to produce sub-optimal solutions achieving a high quality of service (QoS) that satisfies the requirements of users.

In contrast to the works in the literature. In this paper, the resource allocation problem is addressed in steerable laser-based optical wireless networks. We first formulate an optimisation problem that can be solved using MILP with the aim of maximizing the overall SNR of the network. Subsequently, the optimisation problem is reformulated from the RL perspective to avoid the limitations of MILP, while providing sub-optimal solution. To further enhance the overall SNR of the network, beam steering using liquid crystal holograms is considered where more power is focused / steered towards the users. Simulation results show the optimality of the RL model. Moreover, the SINR is significantly improved when beam steering is implemented.

The remainder of this paper is organised as follows: the system model is discussed in Section II. The resource allocation optimisation problem is formulated using MILP and Q-learning in Section III. The simulation configuration and results are presented and discussed in Section IV. Finally, the conclusions are provided in Section V.

## II. SYSTEM MODEL

An OWC system is considered in an empty room with dimensions Width ($x$), Length ($y$), and Height ($z$) as shown in Fig. 1. On the ceiling, $L$ VCSEL arrays are deployed to serve $K$ users uniformly distributed on the receiving plane. Each array is composed of $N$ access points (APs) to serve multiple users simultaneously. It is worth mentioning that each AP comprises multiple VCSELs that jointly serve only one user at a given time to avoid the multi-user interference. An optical line terminal (OLT) that connects all the arrays together is located on the middle of the ceiling with the responsibility of resource management. Each user is equipped with a single-wide field of view (wFOV) receiver in order to collect, filter, and extract the desired data.

The optical channel is composed of Line-of-Sight (LoS) and diffuse (Non-LoS) components. In particular, the LoS component denotes the direct link between the user and transmitter, while the diffuse component is received due to the reflections from the walls and ceiling of the room. Thus, the optical channel between user $k$ and AP $n$ can be expressed as

$$h^{[k,n]} = h_{LoS}^{[k,n]} + h_{diff}(f)e^{-j2\pi f \Delta T}, \quad (1)$$

where $h_{LoS}^{[k,n]}$ denotes the LoS channel, $h_{diff}$ is the diffuse channel and $\Delta T$ is the delay between both components. Notice that, each VCSEL illuminates a confined area limited to a few centimetres. In addition, the implementation of beam steering results in focusing the transmitted power towards the users. Therefore, the diffuse component can be neglected, for the sake of simplicity, where most of the received power is due to the LoS component.

The transmitted power of the VCSEL can be determined based on the beam waist $W_0$, the wavelength $\lambda$ and the distance $d$ between the ceiling and the receiving plane. Therefore, the beam radius of the VCSEL at user $k$ located on the receiving plane is defined as a function of $W_0$, $\lambda$ and $d$. Notice that, the VCSEL has a Gaussian beam profile. Thus, the intensity of the VCSEL can be determined from the radial distance $r$ from the centre of the beam spot and the distance $d$, finding the spatial distribution of the intensity of VCSEL over the transverse plane.

In this work, the performance metric considered is SINR. Two sources of noise are considered. The first source is the thermal noise at the preamplifier of the receiver. The second source is the electrical interference power received from other neighbouring arrays, represented by the dashed lines in Fig. 1. For simplicity, we denote all the types of noise at user $k$ by $\sigma_k$. At this point, each user assigned to an AP receives interference from the neighbouring APs of the same array. The interference is a power received due to data transmission from the neighbouring APs to other users. Thus, the SINR of user $k$ assigned to AP $n$ of array $l$ is given by

$$SINR_{k,n,l} = \frac{P_{k,n,l}}{\sum_{\substack{n' \in N \\ n' \neq n}} P_{k,n',l} + \sigma_k^2} \quad (2)$$

where $P_{k,n,l}$ is the electrical signal power received by user $k$ from AP $n$ of array $l$. Notice that, in our optimisation problem in Section III, we define a constraint that allows each user to be assigned to an exclusive AP in the array. However, the interference among the multiple APs of an array must be managed. Given this point, the bandwidth is divided into four slots, assigning each slot to an AP. Therefore, the coverage area of an array comprises four sectors, allowing a user served by a sector to manage the interference received due to the transmission from the neighbouring sectors as noise. Besides, beam steering using liquid crystal based holograms similar to [3], [11], [12] and [46] is considered to focus the power of each AP towards its assigned user, and therefore, the interference can be minimised, enhancing the performance of the system in terms of the SINR.

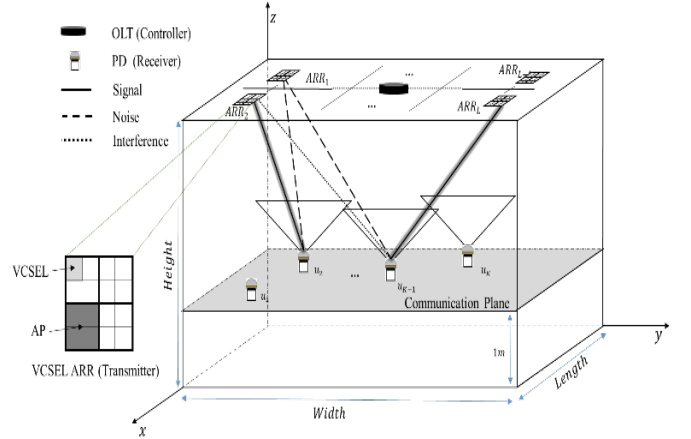

Fig. 1. System Model

## III. PROBLEM FORMULATION

In general, the SINR is an important metric in cellular networks that must be addressed to achieve high performance. In optical wireless networks, users usually experience low SINR due to the large number of overlapping APs required to ensure coverage. In the following, a MILP model is formulated to maximise the SINR by assigning each user exclusively to the best available AP of an array, while managing the received power from other arrays as noise. To provide a reliable and practical solution, the problem is reformulated such that it can be solved using the QL algorithm.

### A. MILP-based Resource Allocation.

In this sub-section, a MILP model is developed to optimise the resource allocation based on maximising the

sum SINR of all users in the room. In this sense, the objective function is represented by

$$\text{maximise} \sum_{k \in K} \sum_{l \in L} \sum_{n \in N} SINR_{k,l,n}. \quad (3)$$

To maximise this objective functions, we define three constraints that play a major role in the optimisation problem. First, each user must be assigned to only one AP of an array, i.e.,

$$\sum_{n \in N} \sum_{l \in L} x_{k,l,n} = 1, \quad \forall\, k \in K. \quad (4)$$

Second, each AP must serve a maximum of one user. It can be can guaranteed using the following constraint

$$\sum_{k \in K} x_{k,l,n} \leq 1, \quad \forall\, n \in N, \forall\, l \in L. \quad (5)$$

To ensure that each user experiences a high SINR that can satisfy the demands in terms of the QoS, a minimum SINR for each user, which is 15.6 dB in this work [35], must be assured by the constraint

$$SINR_{k,l,n} \geq 15.6\, dB\ \forall\, k \in K, l \in L, n \in N. \quad (6)$$

This MILP model was be solved using CPLEX 12.10.

*B. Intelligent Resource Allocation using Q-learning.*

In order to apply the objective function and constraints of the proposed MILP model using reinforcement learning, the optimisation problem is reformulated as a Markov Decision Process (MDP) problem [47]. The MDP problem is used to represent stochastic decision-making problems and can be solved using dynamic programming and reinforcement learning algorithms [37], [38]. Any MDP problem can be represented using five components: agent, environment, state-space $S$, action-space $A$, and reward $R$.

In this context, the environment represents the OWC system that we are aiming to maximise its SINR and the agent is a software located in the control unit that performs the resource allocation decision. Each state $s \in S$ in our environment is defined using a binary vector with length given by the total number of users, $K$, ( $s = \{QoS_1, \dots, QoS_K\}$). If the minimum quality of service required by user $k \in K$ is met, its associated field $QoS_k$ will be equal to 1, otherwise it is 0. That is

$$QoS_k = \begin{cases} 1, & SINR_{k,l,n} \geq 15.6\ dB \\ 0, & \text{otherwise} \end{cases}, \quad (7)$$
$$\forall\, k \in K, l \in L, n \in N$$

Notice that, solutions might be considered only if they lead to a state where all the fields of the users are equal to 1 in order to guarantee the demands of the users. More requirements such as fairness and power consumption can be considered in the future.

The action-space $A$ represents all possible solutions that are considered for the optimisation problem. Each action $a \in A$ is a binary matrix $x$ that indicates the association between an AP of an array and a user. For instance, If user $k \in K$ is assigned to AP $n \in N$ of array $l \in L$, their association variable $x_{k,l,n}$ will be equal to 1, otherwise it will be equal to 0. To reduce the time and memory complexity, pre-processing is done on the action-space to eliminate actions that do not meet the constraints (4) and (5). RL aims to find the optimal policy $\pi^*$ that maximises a certain reward. Since the main objective in our work is to maximise the aggregate SINR, the instantaneous reward function $R$ is represented by

$$R = \sum_{k \in K} SINR_{k,l,n}. \quad (8)$$

In the Q-learning algorithm, the total expected reward from taking an action $a \in A$ in any state $s \in S$ can be measured using Q-values $Q_\pi(s, a)$. These values provide guidance to the controller to decide the best action to take. Initially, all Q-values are set to zero as the agent has no prior knowledge of the network. Therefore, the Q-learning algorithm uses an $\epsilon$-greedy algorithm to balance the exploration/exploitation trade-off. The exploration factor $\epsilon$ is initially set to 1 for the agent to explore the Q-values of new actions. During the learning process, the value of $\epsilon$ gradually decreases and another value $z$ in the range "0" to "1" is randomly chosen. If the value of $z$ is greater than $\epsilon$, the agent will decide to exploit the best learnt Q-value, otherwise, it will explore a new action and update its associated Q-value using the bellman equation in (9).

$$Q_\pi^{new}(s, a) = (1 - \alpha) Q_\pi(s, a) + \alpha [r(s, a) + \gamma \max_a Q_\pi(s', a)], \quad (9)$$

where $\alpha$ is the pre-defined learning rate that can have a value in the range "0" to "1". The purpose of using $\alpha$ is to describe how much did the agent learn about the environment when this Q-value is known. In other words, if $\alpha$ has a large value that approaches "1", the agent will converge quickly while the solution provided will be inaccurate. On the other hand, if $\alpha$ has a very small value that approaches 0, the agent will take a very long time to converge to an accurate optimal solution. Therefore, the value of $\alpha$ must be chosen carefully to find a balance between the learning time and the optimality of the solution. Since our system is continuous, the discount rate $\gamma$ must be defined with a value smaller than "1" in order to allow convergence of the total expected reward. This process is done recursively until the all Q-values within the Q-table converge to an approximate value (reaching the optimal policy $\pi^*$) or the number of iterations exceeds a pre-defined threshold. After the learning process is finished, the optimal policy is chosen by selecting the action that provides the best Q-value $Q^*(s, a)$, i.e.,

$$\pi^* = \max_a Q^*(s, a) \quad (10)$$

IV. SIMULATION SETUP AND RESULTS

The room considered in the simulation contains $L = 4$ VCSEL arrays, each with $N = 4$ APs, on the ceiling serving $K = 4$ users distributed on the communication plane (1m above the floor). Two different user distribution scenarios are considered. 1) Users are uniformly distributed on the communication plane, where each user is located right below one of the arrays. This scenario is defined as the best-case. 2) Users are very close to each other crowded under one of the arrays, which is defined as the worst-case. Both scenarios are tested with and without the deployment of beam steering. All the other parameters such as Room dimensions, transmitter and receiver characteristics, etc., are given in Table 1. It should be noted that the VCSEL power in Table 1 was selected based on the eye safety study in [49].

TABLE I. SIMULATION CONFIGURATION

| Simulation Parameters | |
|---|---|
| Room Dimensions $(x, y, z)$ | $4m \times 4m \times 3m$ |
| *Transmitter Parameters* | |
| Number of access points per array | 4 |
| Number of VCSELs per access point | 4 |
| VCSEL wavelength | $850\ nm$ |
| Total transmitted power of each VCSEL | $5\ mW$ |
| Beam steering angle | 4° |
| Transmitter Locations | Array (1) (1,1,3) <br> Array (2) (1,3,3) <br> Array (3) (3,1,3) <br> Array (4) (3,3,3) |
| *Receiver Parameters* | |
| Photodetector FOV | 40° |
| Photodetector Bandwidth | $5\ GHz$ |
| Noise spectral density | $4.47\ pA/\sqrt{Hz}$ [4] |
| Photodetector area | $55\ mm^2 \times 55\ mm^2$ |
| Responsivity of the receiver | $0.54\ A/W$ |
| Receivers Locations (Scenario 1) | User (1) (1,1,1) <br> User (2) (1,3,1) <br> User (3) (3,1,1) <br> User (4) (3,3,1) |
| Receiver Locations (Scenario 2) | User (1) (3.5,3.5,1) <br> User (2) (3.5,2.5,1) <br> User (3) (2.5,3.5,1) <br> User (4) (2.5,2.5,1) |

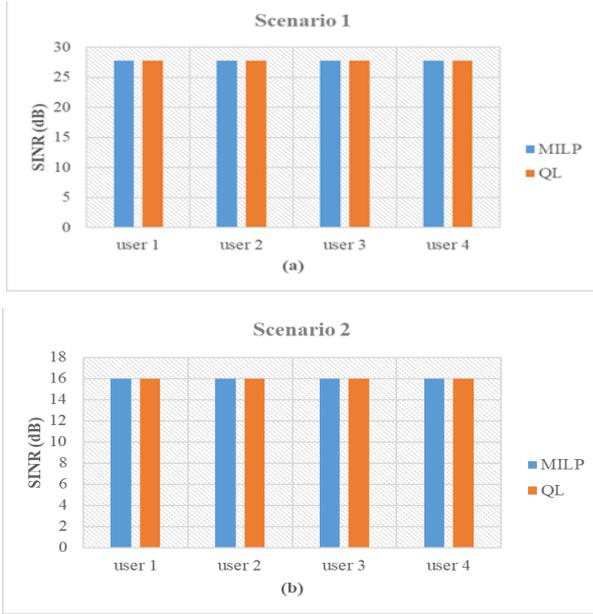

Fig. 2. SINR per user. (a) Scenario 1. (b) Scenario 2.

The optimisation problem is solved using MILP and QL without considering the implementation of beam steering as demonstrated in Figs. 2 and 3. It is shown that the QL algorithm without any prior knowledge of the network achieves optimal solutions significantly close to the output of the MILP model, which requires full knowledge of the environment. Specifically, Fig. 2 shows that the QL algorithm achieves a high SINR close to the MILP model in the scenario where all the users are uniformly distributed on the communication plane. In the worst scenario when the users are located very close to each other below a VCSEL array, it is shown that the users experience a lower SINR regardless of the optimality of the QL algorithm and MILP model due to the fact that all the users are served by an array on different AP generating high noise among them. Furthermore, Table 2 shows that the allocation resulting from the QL algorithm is similar to the allocation of the MILP model in scenario 1. While the solution of the QL algorithm is slightly different from the MILP model in scenario 2. However, it can be seen that the overall SINR of the QL algorithm is similar to the MILP model as shown in Fig. 3.

TABLE II. RESOURCE ALLOCATION RESULTS

| Resource Allocation Results | | | | | | | | |
|---|---|---|---|---|---|---|---|---|
| User # | Scenario 1 | | | | Scenario 2 | | | |
| | MILP | | QL | | MILP | | QL | |
| | Array | AP | Array | AP | Array | AP | Array | AP |
| 1 | 1 | 1 | 1 | 1 | 4 | 3 | 4 | 4 |
| 2 | 2 | 1 | 2 | 1 | 4 | 1 | 4 | 2 |
| 3 | 3 | 1 | 3 | 1 | 4 | 4 | 4 | 1 |
| 4 | 4 | 1 | 4 | 1 | 4 | 2 | 4 | 3 |

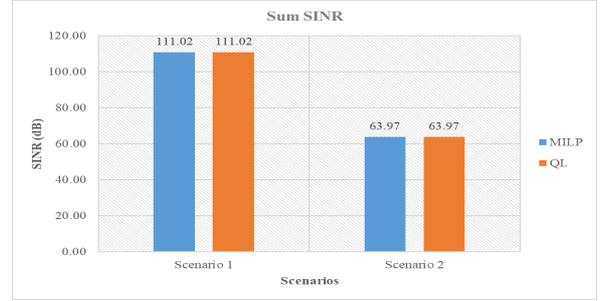

Fig. 3. The sum SINR considering two scenarios 1 and 2.

In Fig. 4, beam steering using liquid crystal based holograms is adopted after assigning the users in order to further enhance the SINR of the network. It can be seen that by steering the power of the APs towards their users within a 4° angle, the SINR of each user is significantly improved for both scenarios compared to APs with free beams.

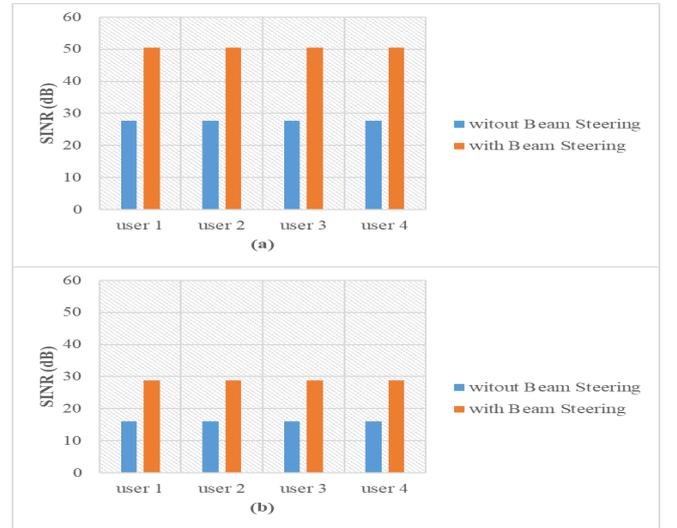

Fig. 4. SINR per user with and without beam steering. (a) Scenario 1. (b) Scenario 2

It is worth pointing out that the QL algorithm can achieve optimal solutions similar to the output of the MILP without the need for prior information. However, the QL algorithm has high computation complexity and memory requirements that must be taken into consideration in future studies. Advanced reinforcement learning techniques such as deep reinforcement learning will be adopted as a promising tool in order to relax the demands in terms of the time and memory. In addition, various optimisation problems with different contexts will be formulated to address resources such as time, frequency, and power to maximise the sum rate of OWC systems.

## CONCLUSIONS

In this paper, a MILP model and QL-based algorithm for resource allocation in a steerable laser-based OWC system is investigated. First, an optimisation problem is formulated as a MILP model. Then, it is reformulated in order to use the QL algorithm as a practical solution. Both techniques aim to assign users to a certain AP within a VCSEL array optimally to achieve the best possible SINR. The results demonstrated that the QL algorithm can achieve solutions with a high SINR similar to the MILP model without prior knowledge of the network. Finally, beam steering using liquid crystal based holograms is implemented to focus the transmitted power towards the users. Therefore, the SINR is improved considerably. In the future, more efficient reinforcement learning techniques with different contexts will be proposed for steerable VCSEL-based systems taking into consideration the complexity and memory requirements.

## ACKNOWLEDGMENTS

This work has been supported in part by the Engineering and Physical Sciences Research Council (EPSRC), in part by the INTERNET project under Grant EP/H040536/1, and in part by the STAR project under Grant EP/K016873/1 and in part by the TOWS project under Grant EP/S016570/1. All data are provided in full in the results section of this paper. The first author would like to acknowledge EPSRC for funding his PhD scholarship.